\documentclass[12pt]{article} 
\usepackage[hyperfootnotes=false]{hyperref}
\usepackage{epsfig}
\usepackage{amsmath}
\usepackage{amssymb}
\usepackage{graphicx}
\usepackage[dvipsnames]{xcolor}
\definecolor{db}{rgb}{0,0,0.7}
\definecolor{dg}{rgb}{0,0.65,0}
\definecolor{pi}{rgb}{1,0.2,0.2}
\usepackage{slashed}
\usepackage{float}
  \newlength{\abstractwidth}
\usepackage{comment}
 \usepackage{mciteplus} 
\usepackage{subcaption}
 \usepackage{bbold}

  \newcommand{\be}{\begin{equation}}
  \newcommand{\bea}{\begin{eqnarray}}
  \newcommand{\eea}{\end{eqnarray}}
  \newcommand{\beq}{\begin{equation}}
  \newcommand{\ee}{\end{equation}}
  \newcommand{\eeq}{\end{equation}}

\def\la{\label}

\def\32{{3 \over 2 } }

  \def\ba{\begin{eqnarray}}
  \def\ea{\end{eqnarray}}

 \def\simleq{\; \raise0.3ex\hbox{$<$\kern-0.75em
      \raise-1.1ex\hbox{$\sim$}}\; }
 \def\simgeq{\; \raise0.3ex\hbox{$>$\kern-0.75em
	\raise-1.1ex\hbox{$\sim$}}\; }

\def\nref#1{(\ref{#1})}

\def\vectork{\vec{\mathbf{k}}}
\def\sk{\vskip .1cm}

\begin{document}

\begin{center}
{\LARGE {\bf Two-Point String Amplitudes}}\\
\vskip 1cm {\large \bf Harold Erbin$^{(a)}$, Juan Maldacena$^{(b)}$ and Dimitri Skliros$^{(a,c)}$}
\vskip 0.2cm {\sl (a) Arnold Sommerfeld Center for Theoretical Physics,\\
Fakult\"at f\"ur Physik, Ludwig-Maximilians-Universit\"at M\"unchen,\\
 Theresienstr. 37, 80333 M\"unchen, Germany}
\vskip 0.08in {\sl (b) Institute for Advanced Study, Princeton, NJ 08540, USA}
\vskip 0.08in {\sl (c) Max-Planck-Institut f\"ur Physik, Werner-Heisenberg-Institut, \\
F\"ohringer Ring 6, 80805 M\"unchen, Germany}

\vskip 0.14in {\tt \small harold.erbin@physik.lmu.de;\,\,\,\,\,\,malda@ias.edu;\,\,\,\,\,\,skliros@mpp.mpg.de}
\end{center}

\bigskip
\vskip -0.06in
\begin{abstract}
We show that the two-point  tree level amplitude in string theory in flat space is given by the standard free particle expression. 
\end{abstract}
 
 \bigskip
 
\section{Introduction}\label{sec:TP}
Consider a string theory with $D$ flat spacetime directions, with Poincar\'e  symmetry. 
The standard lore in string perturbation theory says that the sphere (or tree-level) 2-point amplitude, which we denote by $\mathcal{A}_2 $,
 is zero, $\mathcal{A}_2 =0$.
 \sk
 
The usual argument is that $\mathcal{A}_2 $ vanishes because the residual symmetry group, after fixing two points on the worldsheet, still has infinite volume. Namely, when we fix an operator at zero and the other at infinity the dilatations are still unbroken and have infinite volume. 
 But the situation is subtle because there is also an infinity in the numerator -- 
 the argument of the energy-conservation delta function is automatically zero when the two particles are onshell and the momentum-conservation delta functions are obeyed. 
So in fact we actually end up with an expression of the form:
\begin{equation}\label{eq:A1->1 88}
\mathcal{A}_2 \propto\frac{\infty}{\infty}.
\end{equation}
Consequently, it is also conceivable that the string path integral yields the full connected S matrix (including the interaction-free piece), in which case we should have:
\begin{equation}\label{eq:A1->1=1}
\mathcal{A}_2 = 2k^0(2\pi)^{D-1}\delta^{D-1}(  \vec{\mathbf{k}}'- \vec{\mathbf{k}})
~,~~~~~~~~~~~~k^0 \equiv \sqrt{ m^2 + {\vec{\mathbf{k}}}^2 },
\end{equation}
where we consider momentum eigenstates and adopt covariant normalization. 
In cases where the mass eigenstates are degenerate there is an extra  Kronecker delta, $\delta_{j',j}$, with $j$ and $j'$  running over an orthonormal basis. We suppress this simple factor from now on. 
\sk

Notice that if we know that the result is non-zero, we can determine its normalization 
via a unitarity argument. Namely, we should require that:
\begin{equation}\label{eq:A=AA}
\mathcal{A}_2 
(\vectork',\vectork'')=
\int \frac{d^{D-1}\mathbf{k}}{(2\pi)^{D-1}}\frac{1}{2k^0}\,
\mathcal{A}_2 (\vectork',\vectork)\mathcal{A}_2 (\vectork,\vectork'').
\end{equation}
Using (\ref{eq:A=AA})  we learn that (\ref{eq:A1->1=1}) must be the correct answer, {\it if} we can show that $\mathcal{A}_2 \neq0$. 
\sk

We will  argue that we indeed get 
(\ref{eq:A1->1=1}),   after properly defining the ratio in (\ref{eq:A1->1 88}). We first present a qualitative argument and then a more formal argument using the Fadeev-Popov trick.  

\section{The two-point amplitude}\label{sec:PD}
 
 \subsection{Open strings } 
 
Let us consider open strings first. Our goal is to define the ratio in \nref{eq:A1->1 88} in a more precise way.  We find it useful to work in conformal gauge, and to choose   
 coordinates $\tau, \sigma$ where the worldsheet metric is 
 $ds^2 = d\tau^2 + d\sigma^2$, with $\sigma \in [ 0, \pi ] $. Then the integrated vertex operators include only an integral along the boundary $W = \int d\tau V $. And we divide by the volume of ${\rm PSL}(2,\mathbf{R})$. 
\sk
 
 When we consider the two-point function we can fix two points, one at zero and the other at infinity. In order to derive a formal expression for the volume of the residual group, let us consider the three-point function. 
We can first fix two of the three points at $\tau = \pm \infty$.    We then integrate over the third point by inserting an integrated vertex operator, $W_2 = \int   d \tau   V_2(\tau ,0)$, and after dividing by the volume of the residual group we get \cite{Polchinski_v1},
\be \la{Thpt}
\mathcal{A}_3 = g_o^3C_{D_2}\big\langle V_1(-\infty) V_2(0) V_3(\infty) \big\rangle_{D_2} = 
 g_o^3C_{D_2}  \int   { d\tau \over {\rm Vol(Res) } }  \big\langle V_1(-\infty) V_2(\tau) V_3(\infty) \big\rangle_{D_2} 
\ee
where all operators are on the $\sigma=0$ boundary and 
 $C_{D_2}$ is the partition function of all worldsheet fields with an appropriate power of the string coupling determined by unitarity as in \cite{Polchinski_v1}.  The first expression in \nref{Thpt} is just the usual expression for the three-point amplitude. By noticing that the integrand is independent of $\tau$ we 
conclude that 
the volume of the residual conformal group (after fixing two points) is the formal expression
\be \la{OpenM}
{\rm Vol(Res)} = \int_{-\infty}^\infty  d\tau.
\ee

We now consider the matter path integral as follows. First we do the integral over all the spatial dimensions, which  produces the usual delta function,
\be
(2 \pi)^{D-1} \delta^{D-1} ( \vectork_1 - \vectork_2),
\ee
This, together with the mass shell condition automatically implies that 
$k^0_1 = k^0_2 $. So the partition function does
not depend on $x^0$, the zero mode of $X^0$. Here we are thinking of  
$X^0$   as Euclidean, but with $k^0$ as in \nref{eq:A1->1=1} and the vertex operators 
containing $e^{ \pm k^0 X^0}$.  
\sk
 
Using (\ref{OpenM}) the 2-point amplitude takes the form:
\be  \la{TwoPt}
\mathcal{A}_2 = \left( {\int d x^0 \over \int d \tau } \right) g_o^2C_{D_2}   ( 2\pi)^{D-1} \delta^{D-1} ( \vectork_1 - \vectork_2).
\ee

The two integrals in the first factor diverge and we need to compute their ratio. 
We can consider the following classical solution for $X^0$,
\be \la{norm}
X^0 = 2 \alpha' k^0 \tau  +  x^0.
\ee
This can be viewed as the classical solution we expand around in order to compute the path integral.  A feature of \nref{norm} is that it spontaneously breaks the translation symmetry in $\tau$. Of course, this symmetry is restored once we integrate over $x^0$,  the $X^0$ zero mode. 
We now see that a shift of   $x^0$  can be absorbed by a shift of $\tau$,
\be
x^0 \to x^0 + \epsilon ~,~~~~~~~\tau \to \tau - { \epsilon \over 2 \alpha' k^0 } .
\ee
This represents a connection between translations in $X^0$ and translations in $\tau$, and we interpret this as setting  the relative normalization of the two symmetries. 
The idea is that this implies that the ratio of integrals in \nref{TwoPt} can be written as,
\be \la{openfact}
\left( {\int d x^0 \over \int d \tau } \right) = 2 \alpha' k^0.
\ee
 
Substituting (\ref{openfact}) into (\ref{TwoPt}) leads to the final expression for the two-point function,
\be \la{FinAn}
\mathcal{A}_2 = { 2 k^0 } (2 \pi)^{D-1} \delta^{D-1}( \vectork_1 - \vectork_2) 
\left( \alpha' g_o^2 C_{D_2} \right).
\ee
The factor in the last parenthesis is equal to one, see formula  (6.4.14) \cite{Polchinski_v1}. This factor  was determined by factorizing the poles of the four-point amplitude into a product of three-point amplitudes. 
Thus, we have reproduced (\ref{eq:A1->1=1}).  

\subsection{Closed Strings } \label{sec:CS}
 
Let us also mention some of the changes in the case of closed strings. Now we have the periodic identification, $\sigma \sim \sigma + 2 \pi$, and the volume of the residual group is:
\be\label{eq:volres}
{\rm Vol(Res)}  = 2 \int_{-\infty}^\infty  d\tau \int_0^{2\pi} d\sigma,
\ee
the normalization of which is fixed by considering the 3-point function with the points fixed as usual with a product of three $V_i$. The factor of 2 is consistent with Polchinski's normalization for the measure, 
$d^2z = 2 d\tau d\sigma$ if $z = \tau + i \sigma$ (see 2.1.7 in \cite{Polchinski_v1}), and is also consistent with equation 6.6.7 in \cite{Polchinski_v1}, which is used to normalize the sphere amplitude.\footnote{By looking at equation 6.4.10 and  6.4.11 in \cite{Polchinski_v1} we see that for open strings Polchinski is defining the measure as in \nref{OpenM}.}
\sk
 
Since the worldsheet of the closed string has the periodicity, $ \sigma \sim  \sigma + 2 \pi $, equation \nref{norm} is replaced by, 
\be
  X^0 = \alpha' k^0 \tau + x^0,
\ee
Proceeding as in the open string, this then leads to the following relation between shifts of $\tau$ and $X^0$ leading to,
\be
\left( { \int dx^0 \over 2 \int d\tau d\sigma } \right) = { 1 \over 4 \pi } \alpha' k^0.
\ee
 
So the final expression of the two-point function is:
\be
\mathcal{A}_2 = 2 k^0 (2 \pi)^{D-1} \delta(\vectork_1 - \vectork_2) \left( g_c^2 {\alpha' \over 8 \pi } C_{S^2} \right).
\ee
The last factor is precisely equal to 1, see formula 6.6.8 of \cite{Polchinski_v1}.  We again get (\ref{eq:A1->1=1}).
\sk
 
For the NS-NS sector of the superstring there is an identical story when 
we put both operators in the $-1$ picture. For other sectors it should also work similarly due to spacetime supersymmetry.  

\section{Derivation using the Fadeev Popov trick } 

\subsection{Short review of the Fadeev Popov trick }

  We consider an integral over some variables $x^a$ which is invariant under a lie group $G$. We want to give meaning to the expression:
  \be \la{Zinte}
  Z= \int { Dx  \over {\rm Vol}(G) } I(x).
  \ee
  This is non-trivial when $G$ is a non-compact group. 
  Suppose that $d$ is the dimension of $G$. Then we choose $d$ ``gauge fixing functions'' $f_i(x)$, $i=1, \dots, d$, which are {\it not} invariant under $G$. We will assume that the zeros of these functions intersect the orbits of $G$ just once. 
  We then define $\Delta$ through the equation,
  \be \la{OneInFP}
   1 =  \Delta(x)    \int Dg \prod_{i=1}^d \delta( f_i(x^g) ), 
  \ee
  where the integral is over the group and $x^g$ is the result of acting by the group element $g$ on $x$. Note that $\Delta(x)$ is invariant under $g$. 
  We now insert 1, written as in \nref{OneInFP}, into \nref{Zinte}. Then we use the invariance of $I(x) $  and 
  $\Delta(x)$ under $g$ to replace 
  $x \to x^g$ in their arguments as well as in the measure. 
  Finally, we relabel the integration variables $x^g \to x$, and use that formally ${\rm Vol }(G) = \int D g $ to cancel the integral over $G$ in the numerator with the group volume in the denominator.  
\sk
  
  In the usual applications to tree level string theory we first do the integral over the worldsheet fields. We are then left with integrals over the positions of the vertex operators. These positions are the variables $x^a$ in \nref{Zinte}.  For open strings, we should divide by the volume of ${\rm PSL}(2,\mathbf{R})$. This is done by picking three gauge fixing functions which fix three vertex operators at positions $z_1^0$, $z_2^0$, $z_3^0$ along the boundary of the worldsheet.  In this particular case we find that,
  \be
  \Delta = z_{12}^0 z_{13}^0 z_{23}^0.
  \ee
  Here we imagine that the worldsheet is the upper half plane and that  $z_i$ and $z_i^0$ lie along the real axis. This expression can also be viewed as the result of inserting ghost fields. 

\subsection{Using the Fadeev Popov trick for the open string two-point amplitude}

In the case of the two-point amplitude we want to include the two operator positions as well as the worldsheet field $X^0$ among the variables 
$x^a$ in \nref{Zinte}. We can then choose the three gauge-fixing functions,
  \be \la{ChFn} 
   f_1 =  z_1 -z_1^0 ~,~~~~~~~f_2 = z_2 -z_2^0 ~,~~~~~~f_3 = X^0(z_3^0,\bar z_3^0),
   \ee
and find that $\Delta$ is given by:
   \be
\Delta = {z_{12}^0 z_{31}^0 z_{32}^0 } \partial_z X^0 + {\bar z_{12}^0  \bar z_{31}^0 \bar z_{32}^0  } \bar \partial_{\bar z} X^0 ~,~~~~~z^0_{1,2} = \bar z_{1,2}^0.
  \ee
  Note that both $z^0_1$ and $z_2^0$ are real because they are on the worldsheet boundary. 
\sk
    
Inserting $\Delta$ into the path integral requires the expectation value,
   \be \la{ExpVa}
 \langle \partial X^0 V_1 V_2 \rangle =    i { \alpha' k^0 } \left( { z_{12}  \over (z - z_1) ( z- z_2 ) } \right) \langle V_1 V_2\rangle,
  \ee
  and a similar one involving $\bar \partial_{\bar z} X$ where we used that $z_1$, $z_2$ are on the boundary. The $z$ dependence in \nref{ExpVa} is fixed by conformal symmetry. The overall factor is fixed by noticing that the contour integral of $\partial X$ around each vertex operator computes the spacetime  energy $k^0$.  
    \sk
     
 We then conclude that,
 \be
 \langle \Delta \delta(X^0(z_3,\bar z_3) ) V_1 V_2 \rangle = { 2 \alpha' k^0 } (z^0_{12})^2 \langle V_1 V_2 \rangle',
 \ee
where the prime means that we do not integrate over the zero mode of the time direction. Note that the final result is independent of
$z_3^0$ and $\bar z_3^0$ which appeared in \nref{ChFn}.  Of course, the $(z^0_{12})^2$ cancels a similar factor in the denominator of $\langle V_1 V_2 \rangle'$.  
It is also possible to rewrite $\Delta$ in terms of ghosts but it is not particularly illuminating.
\sk

Note that the expression for the amplitude involves a delta function along the time direction. This is similar to the expression for the Klein Gordon norm which involves an integral over space at a fixed value of the time coordinate. 
\sk
  
  In the closed-string case we can perform similar gauge fixing. We can fix the locations of the two vertex operators and then introduce a delta function involving the $X^0$ coordinate. For example, if we choose a metric on $S^2$ in spherical coordinates we could insert a delta function for the average value of 
  $X^0$ on the sphere.  Again, one can check that this produces the right factor to give the final result \nref{eq:A1->1=1}.  We  omit the details.  
 
\section{Discussion} 

We have discussed how to obtain a non-zero value for the two-point function in tree-level string theory. 
\sk

Non-zero values for two-point functions are also important for AdS correlators, and a similar cancellation was discussed in \cite{Kutasov:1999xu,Giveon:2001up}. 
\sk

It would be interesting to understand how to obtain non-zero values for lower-point functions. In particular, the zero-point function, or onshell action, for non-trivial backgrounds, such as a black hole background, for example. Such low-point functions have been obtained in non-critical string theory \cite{Zamolodchikov:1982vx}, see   \cite{Seiberg:1990eb, Dorn:1994xn}  for a review.

\section*{Acknowledgements}

  We would like to thank Subhroneel Chakrabarti, Raghu Mahajan and Ashoke Sen for insightful discussions and remarks. 
\sk
  
The work of H.E.\ is conducted under a Carl Friedrich von Siemens Research Fellowship of the Alexander von Humboldt Foundation for postdoctoral researchers. J.M. is supported in part by U.S. Department of Energy grant
de-sc0009988 and by the Simons Foundation grant 385600. D.S.~is supported by the Max Planck Institute for Physics in Munich.  We also thank the Galileo Galilei Institute for Theoretical Physics and INFN for hospitality and partial support during the workshop "String Theory from a worldsheet perspective" where part of this work was carried out. 

\bibliographystyle{JHEP}


\end{document}